\documentclass[a4paper,12pt]{article}
\usepackage{jcappub}
\usepackage[symbol]{footmisc}
\usepackage{amsmath,amssymb}
\usepackage{revsymb}
\usepackage{orcidlink}
\newcommand{\nc}{\newcommand*} 

\def\({\left(}
\def\){\right)}
\def\[{\left[}
\def\]{\right]}

\nc{\Eq}[1]{Eq.~\eqref{#1}}
\nc{\Fig}[1]{Fig.~\ref{#1}}     
\nc{\Table}[1]{Table~\ref{#1}}  
\nc{\Sec}[1]{Sec.~\ref{#1}}     

\nc{\Msun}{M_\odot}             
\nc{\fpbh}{f_{\mathrm{PBH}}}    
\nc{\mR}{\mathcal{R}} 
\nc{\seq}{\sigma_{\mathrm{eq}}}
\nc{\ogw}{\Omega_{\mathrm{GW}}}
\nc{\gpcyr}{\mathrm{Gpc}^{-3}\,\mathrm{yr}^{-1}}
\nc{\lvc}{LIGO/Virgo} 
\nc{\SNR}{\mathrm{SNR}} 
\nc{\mmin}{{m_{\mathrm{min}}}}
\nc{\mmax}{{m_{\mathrm{max}}}}
\nc{\Mmin}{{M_{\mathrm{min}}}}
\nc{\fmin}{{f_{\mathrm{min}}}}
\nc{\VT}{\mathrm{VT}}
\nc{\rhoGW}{\rho_{\mathrm{GW}}}
\nc{\vth}{\vec{\theta}}
\nc{\vd}{\vec{d}}
\nc{\vla}{\vec{\lambda}}
\nc{\Nobs}{N_{\mathrm{obs}}}
\nc{\av}[1]{\langle #1 \rangle} 
\nc{\km}{\mathrm{km}}
\nc{\Mpc}{\mathrm{Mpc}}
\nc{\Tobs}{T_{\mathrm{obs}}}
\nc{\Ntemp}{N_{\mathrm{temp}}}

\nc{\addref}{[\textcolor{red}{add ref}] } 
\nc{\eg}{\textit{e.g.~}}
\nc{\app}{\approx}
\nc{\hf}{\frac{1}{2}}
\nc{\discuss}{\textcolor{red}{Add discussion here!}}

\nc{\mpbh}{m_{\rm{pbh}}}
\nc{\cR}{\mathcal{R}}
\nc{\mU}{{\mathcal{U}}}
\nc{\Mc}{{M_\mathrm{c}}}
\nc{\Mf}{{M_\mathrm{f}}}
\nc{\red}[1]{\textcolor{red}{#1}}
\nc{\yellow}[1]{\textcolor{yellow}{#1}}
\nc{\green}[1]{\textcolor{green}{#1}}
\nc{\blue}[1]{\textcolor{blue}{#1}}
\nc{\fnl}{F_{\mathrm{NL}}}
\nc{\gnl}{G_{\mathrm{NL}}}
\nc{\MG}{\mathcal{M}_{\mathrm{G}}}
\nc{\MNG}{\mathcal{M}_{\mathrm{NG}}}
\newcommand{\papertitle}{GW230529\_181500: A Potential Primordial Binary Black Hole Merger\\ in the Mass Gap}

\begin{document}
	
\title{\papertitle} 


\author[a,b,c,d]{Qing-Guo Huang\orcidlink{0000-0003-1584-345X}}
\author[e]{Chen Yuan\orcidlink{0000-0001-8560-5487}}
\author[f,g,*]{Zu-Cheng Chen\orcidlink{0000-0001-7016-9934}}
\author[h,i,*]{Lang~Liu\orcidlink{0000-0002-0297-9633},\note{Corresponding author.}}

\affiliation[a]{Institute for Frontiers in Astronomy and Astrophysics, Beijing Normal University, Beijing 102206, China}
\affiliation[b]{CAS Key Laboratory of Theoretical Physics, Institute of Theoretical Physics, Chinese Academy of Sciences, Beijing 100190, China}
\affiliation[c]{School of Physical Sciences, University of Chinese Academy of Sciences, No. 19A Yuquan Road, Beijing 100049, China}
\affiliation[d]{School of Fundamental Physics and Mathematical Sciences, Hangzhou Institute for Advanced Study, UCAS, Hangzhou 310024, China}
\affiliation[e]{CENTRA, Departamento de Física, Instituto Superior Técnico - IST, Universidade de Lisboa - UL, Avenida Rovisco Pais 1, 1049-001 Lisboa, Portugal}
\affiliation[f]{Department of Physics and Synergetic Innovation Center for Quantum Effects and Applications, Hunan Normal University, Changsha, Hunan 410081, China}
\affiliation[g]{Institute of Interdisciplinary Studies, Hunan Normal University, Changsha, Hunan 410081, China}
\affiliation[h]{Department of Astronomy, Beijing Normal University, Beijing 100875, China}
\affiliation[i]{Advanced Institute of Natural Sciences, Beijing Normal University, Zhuhai 519087, China}

\emailAdd{huangqg@itp.ac.cn}
\emailAdd{chenyuan@tecnico.ulisboa.pt}
\emailAdd{zuchengchen@hunnu.edu.cn}
\emailAdd{liulang@bnu.edu.cn}
	
\abstract{
During the fourth observing run of the LIGO-Virgo-KAGRA detector network, the LIGO Livingston observatory detected a coalescing compact binary, GW230529\_181500, with component masses of 2.5-4.5\,$M_\odot$ and 1.2-2.0\,$M_\odot$ at the $90\%$ credible level. The gravitational-wave data alone is insufficient to determine whether the components are neutron stars or black holes. In this paper, we propose that GW230529\_181500 originated from the merger of two primordial black holes (PBHs). We estimate a merger rate of $5.0^{+47.0}_{-4.9} \mathrm{Gpc}^{-3}\,\mathrm{yr}^{-1}$ for compact binary coalescences with properties similar to GW230529\_181500. Assuming the source is a PBH-PBH merger, GW230529\_181500-like events lead to approximately $1.7^{+36.2}_{-1.5} \times 10^{-3}$ of the dark matter in the form of PBHs. The required abundance of PBHs to explain this event is consistent with existing upper limits derived from microlensing, cosmic microwave background observations and the null detection of gravitational-wave background by LIGO-Virgo-KAGRA.
}
	
\maketitle
\section{Introduction} 	
The existence of a ``lower mass gap'' in the mass distribution of compact objects, spanning approximately $3\,\Msun$ to $5\,\Msun$, has been proposed since the late 1990s~\cite{Bailyn:1997xt,Ozel:2010su,Farr:2010tu,Fishbach:2020ryj}. This gap is thought to separate the heaviest neutron stars (NSs) from the lightest stellar-mass astrophysical black holes (ABHs). However, recent observations of binary systems through electromagnetic and gravitational waves (GWs) have uncovered potential candidates with components falling within this mass gap. A notable example is GW190814, where the secondary object's mass was estimated with high confidence to be between $2.50\,\Msun$ and $2.67\,\Msun$~\cite{LIGOScientific:2020zkf}, exceeding the heaviest known NS at the time but below the expected ABH masses. 

The recent detection of GW230529\_181500 (hereafter referred to as GW230529) by the LIGO-Virgo-KAGRA (LVK) Collaboration presents an even more compelling case~\cite{LIGOScientific:2024elc}. With the primary mass estimated at around $3.6\,\Msun$, GW230529 is the first binary candidate to have its primary component firmly within the lower mass gap. Based on the current understanding of NS and ABH populations, the primary mass is consistent with a black hole smaller than $5\,\Msun$ at $99\%$ probability. 
However, the possibility that the primary component is an unusually heavy NS cannot be entirely dismissed, with the probability potentially reaching a few percents under certain assumptions, taking into account current knowledge from nuclear physics theory, experiments, and astrophysical source populations~\cite{LIGOScientific:2024elc}. The tidal deformability of the secondary object in GW230529 remains unconstrained, and the constraints on the primary object's tidal deformability are consistent with both a black hole and an NS in this mass range, making it difficult to conclusively determine the nature of the compact objects involved in this binary system based solely on tidal effects~\cite{LIGOScientific:2024elc}.

The uncertain nature of the compact objects in GW230529 opens up intriguing possibilities. One such possibility is that the binary components are primordial black holes (PBHs), a distinct class of black holes that could populate the Universe alongside ABHs, which could theoretically populate the lower mass gap. PBHs are hypothesized to have formed in the early Universe through the gravitational collapse of overdense regions~\cite{Zeldovich:1967lct,Hawking:1971ei,Carr:1974nx}. Unlike stellar-mass ABHs that originate from stellar collapse, one of the most striking features of PBHs is their potential to exist across an extensive range of masses, including those within the lower mass gap and below. The study of PBHs has far-reaching implications for our understanding of the Universe~\cite{Chen:2018rzo,Wang:2019kaf,Liu:2020cds,DeLuca:2020sae,Vaskonen:2020lbd,DeLuca:2020agl,Domenech:2020ers,Cai:2021wzd,Yuan:2021qgz,Liu:2021jnw,Liu:2022wtq,Inomata:2022yte,Meng:2022low,Wang:2024vfv,Yuan:2024yyo,Chen:2024dxh,Hai-LongHuang:2023atg}. Besides being a compelling candidate for dark matter~\cite{Sasaki:2018dmp, Carr:2020gox, Carr:2020xqk}, PBHs could contribute to the GW events observed by LVK collaboration~\cite{Bird:2016dcv,Sasaki:2016jop}, offering a unique probe into their properties and abundance. Furthermore, PBHs might serve as the seeds for the formation of galaxies and supermassive black holes~\cite{Bean:2002kx,Kawasaki:2012kn,Nakama:2017xvq,Carr:2018rid}, potentially reshaping our understanding of cosmic structure formation. 
Moreover, the companion to the eccentric binary millisecond pulsar, PSR J0514$-$4002E~\cite{Barr:2024wwl}, can potentially be a PBH~\cite{Chen:2024joj}.

In this paper, we investigate whether the merger rate of GW230529 is compatible with the existing upper limits on PBH abundance, thereby testing the hypothesis that this event originated from a PBH binary coalescence. Our analysis aims to shed light on the potential primordial nature of the compact objects involved in GW230529.
The rest of this paper is structured as follows. In Section~\ref{Merger-rate}, we provide a comprehensive overview of the merger rate of PBH binaries, laying the foundation for our investigation. In Section~\ref{Methodology}, we present a detailed description of the data analysis methodology employed in this study. Lastly, we summarise our conclusions and offer a discussion of the implications in Section~\ref{Result}.

\section{Merger rate of PBH binaries} 
\label{Merger-rate}

PBHs can form binaries through several formation channels~\cite{Nakamura:1997sm,Bird:2016dcv,Sasaki:2016jop,Stasenko:2024pzd}. In this study, we focus on the formation channel of PBH binaries in the early Universe~\cite{Nakamura:1997sm}, 
which is known to make a dominant contribution to the PBH merger rate \cite{Ali-Haimoud:2017rtz}. 
This approach neglects binary formation mechanisms in the late Universe, such as dynamical captures and three-body interactions~\cite{Kritos:2020wcl}. 

We assume that PBHs are initially randomly distributed, following a spatial Poisson distribution in the early Universe, a condition that holds when they decouple from the cosmic background evolution~\cite{Nakamura:1997sm,Sasaki:2016jop,Ali-Haimoud:2017rtz}. Due to the gravitational torque exerted by other PBHs, pairs of PBHs acquire angular momentum, ultimately leading to the formation of a PBH binary upon decoupling from cosmic expansion. The binary then undergoes gravitational radiation emission, potentially resulting in a merger GW event detectable by GW detectors

The merger rate per unit volume at cosmic time $t$ for PBHs within mass intervals of $(m_1,m_1+dm_1)$ and $(m_2,m_2+dm_2)$ is defined as $\mathcal{R}(t,m_1,m_2)dm_{1}dm_{2}$ in units of $\rm Gpc^{-3}~yr^{-1}$. The merger rate density of a PBH binary at cosmic time $t$ is expressed as~\cite{Chen:2018czv,Raidal:2018bbj,Liu:2018ess,Hutsi:2020sol}
\begin{equation}\label{cR}
\begin{split}    
\mathcal{R}(t, m_1, m_2) &= \frac{1.6\times 10^{6}}{\mathrm{Gpc^3\,yr}} \left(\frac{t}{t_0}\right)^{-\frac{34}{37}} f_\mathrm{PBH}^\frac{53}{37} \eta^{-\frac{34}{37}} \left(\frac{M}{M_\odot}\right)^{-\frac{32}{37}} P(m_1) P(m_2) S[P(m), f_\mathrm{PBH}, M],
\end{split}
\end{equation}
where $t_0$ is the present cosmic time, $M=m_1 + m_2$ is the total mass, $\eta\equiv m_1 m_2/M^2$, $f_\mathrm{PBH}=\Omega_{\mathrm{pbh}}/\Omega_{\mathrm{m}}$ is the total fraction of dark matter in PBHs, and the suppression factor $S[P(m), f_\mathrm{PBH}, M]$ which is given by Ref.~\cite{Hutsi:2020sol} takes into account the influence of the surrounding smooth matter component on binary formation and the potential disruption of the binary system by PBHs.
The contribution of hierarchical mergers~\cite{Liu:2019rnx} has been neglected, given its subdominance as constrained by the GW observations~\cite{Wu:2020drm,Liu:2022iuf}.
Additionally, the total merger rate can be obtained by integrating the component masses as
\begin{equation}\label{Rt}
R(t) = \int \mathcal{R}(t, m_1, m_2)\, \mathrm{d} m_1\, \mathrm{d} m_2.
\end{equation} 
The cosmic time $t$ and redshift $z$ are related by
\begin{equation}
t(z) = \int_{z}^{\infty} \frac{dz'}{H(z') (1+z')},
\end{equation}
where $H(z)$ is the Hubble parameter. Contributions from radiation and neutrinos are neglected, given the limited sensitivity of current ground-based GW detectors to a small redshift range. When converting between luminosity distances, times, and redshifts, we adopt the best-fit cosmological model of Planck 2018~\cite{Planck:2018vyg}.

To constrain the PBH scenario, one must place some a priori constraints on the form of the PBH mass function, $P(m)$. The form of the mass function is sensitive to the details of PBH formation. We use a Gaussian PBH mass function that is defined by~\cite{Dolgov:1992pu}
\begin{equation}\label{Pm}
P(m) = \frac{1}{\sqrt{2\pi} \sigma} \exp \left(-\frac{\left(m-M_c\right)^2}{2\sigma^2}\right),
\end{equation}
where $M_c$ is the median mass and $\sigma$ characterizes the width of the mass distribution.
{A PBH population with a Gaussian mass function provides a good approximation of the enhanced primordial power spectrum \cite{Gow:2020cou}. Such enhancement can be generated by scenarios such as ultra-slow-roll inflation~\cite{Ragavendra:2020sop,Pattison:2021oen} or multi-field inflation \cite{Palma:2020ejf,Geller:2022nkr}, which generate the necessary enhancement in the power spectrum to form PBHs. These scenarios deviate from the standard slow-roll approximation and require at least three parameters, $\{f_\mathrm{PBH}, M_c, \sigma\}$, to model the binary merger rate.}

\section{Methodology}\label{Methodology}

The merger rate density, $\mathcal{R}(\theta|\Lambda)$, as defined in \Eq{cR}, is expressed in the source frame. However, to perform the analysis, it is necessary to transform the merger rate density into the detector frame, $\mathcal{R}_{\mathrm{pop}}(\theta|\Lambda)$, by 
\begin{equation}\label{Rpop}
\mathcal{R}_{\mathrm{pop}}(\theta|\Lambda) = \frac{1}{1+z} \frac{dV_\mathrm{c}}{dz} \mathcal{R}(\theta|\Lambda),
\end{equation}
where $z$ is the cosmological redshift, $dV_\mathrm{c}/dz$ represents the differential comoving volume, and $\theta\equiv \{z, m_1, m_2\}$ constitutes the parameters that defining the GW event. In this analysis, we concentrate on the distributions of redshift and mass, while neglecting the spin distribution. 
Moreover, the set of parameters $\Lambda\equiv \{\fpbh, M_c, \sigma\}$ describes the PBH population, and the factor $1/(1 + z)$ in \Eq{Rpop} accounts for the time dilation between the source and detector frames.

We model the number of GW events as an inhomogeneous Poisson process, given the observed data, $\textbf{d}$, of {single observation} GW230529. {While a single observation may not provide strong constraints on the PBH model parameters, it can still offer valuable insights into the consistency of GW230529-like events with a PBH scenario. By focusing on the specific properties of GW230529 and their compatibility with the predictions of the PBH model, we can assess the plausibility of a primordial origin for this event. This single-event analysis serves as a starting point for investigating the potential role of PBHs in explaining compact objects in the lower mass gap, a region where the formation of ABHs and NSs is challenging to explain with current stellar evolution models.} The likelihood function reads~\cite{Loredo:2004nn,Thrane:2018qnx,Mandel:2018mve}
\begin{equation}
\label{L1}
	\mathcal{L}(\textbf{d}|\Lambda) \propto T_{\mathrm{obs}}\,  e^{-N_{\mathrm{exp}}(\Lambda)}  \int \mathcal{L} (\textbf{d}| \theta)\, \mathcal{R}_{\mathrm{pop}}(\theta|\Lambda)\, d \theta,
\end{equation}
where $N_{\exp}(\Lambda) \equiv \xi(\Lambda) T_{\mathrm{obs}}$ represents the Poisson probability of observing the expected number of detections over the observation timespan $T_{\mathrm{obs}}$, and $\mathcal{L} (\textbf{d}|\theta)$ is the likelihood for GW230529, which can be derived from its posterior by reweighing with the prior on $\theta$. 
In this work, we use the ``Combined\_PHM\_highSpin" posteriors as released by LVK~\cite{LIGOScientific:2024elc}.
Furthermore, $\xi(\Lambda)$ accounts for the selection biases introduced by the detector's sensitivity
\begin{equation}\label{xi}
\xi(\Lambda) = \int P_{\mathrm{det}}(\theta)\, \mathcal{R}_{\mathrm{pop}}(\theta|\Lambda)\, \mathrm{d} \theta,
\end{equation}
where $0 < P_{\mathrm{det}}(\theta) < 1$ captures the detection probability~\cite{OShaughnessy:2009szr}, a function of the source parameters $\theta$.
The estimation of $\xi(\Lambda)$ is performed using simulated injections~\cite{ligo_scientific_collaboration_and_virgo_2021_5546676}, where a Monte Carlo integral over found injections~\cite{KAGRA:2021duu} is used to approximate $\xi(\Lambda)$ as \begin{equation}
	\xi(\Lambda) \approx \frac{1}{N_{\mathrm{inj}}} \sum_{j=1}^{N_{\text {found }}} \frac{\mathcal{R}_{\mathrm{pop}}(\theta_{j} | \Lambda)}{p_{\mathrm{draw}}(\theta_j)},
\end{equation}
with $N_{\text{inj}}$ representing the total number of injections, $N_{\text{found}}$ denoting the count of successfully detected injections, and $p_{\mathrm{draw}}$ being the probability density function from which the injections are drawn. In this work, we follow the approach in~\cite{LIGOScientific:2024elc} and use the combined injection sets from the first three observing runs, effectively assuming that GW230529 occurred at the end of O3. While this assumption introduces a small bias in our inferred estimates of the merger rate in the mass gap by not accounting for the extra time-volume provided by the first two weeks of O4, the effect is considered negligible due to the short duration and similar detector sensitivity between the start of O4a and O3~\cite{LIGOScientific:2024elc}.

Using the posterior samples from GW230529, which are available at~\cite{ligo_scientific_collaboration_2024_10845779}, we calculate the hyper-likelihood~\eqref{L1} 
\begin{equation}\label{L2}
\mathcal{L}(\textbf{d}|\Lambda) \propto T_{\mathrm{obs}}\,  e^{-N_{\mathrm{exp}}(\Lambda)} \left\langle \frac{\mathcal{R}_{\mathrm{pop}}(\theta|\Lambda)}{\pi(\theta)} \right\rangle,
\end{equation}
where $\langle\cdots\rangle$ represents the weighted average over posterior samples of $\theta$, and $\pi(\theta)$ refers to the priors on source parameters used to construct the posterior of GW230529. To estimate the likelihood function \eqref{L2}, we incorporate the PBH population distribution \eqref{cR} into the \texttt{ICAROGW}~\cite{Mastrogiovanni:2021wsd} package and employ the \texttt{dynesty}~\cite{Speagle:2019ivv} sampler, which is called from \texttt{Bilby}~\cite{Ashton:2018jfp,Romero-Shaw:2020owr} package, to sample over the parameter space.

\section{Result and discussion}\label{Result}

\begin{figure}[t]
    \centering
    \includegraphics[width=\linewidth]{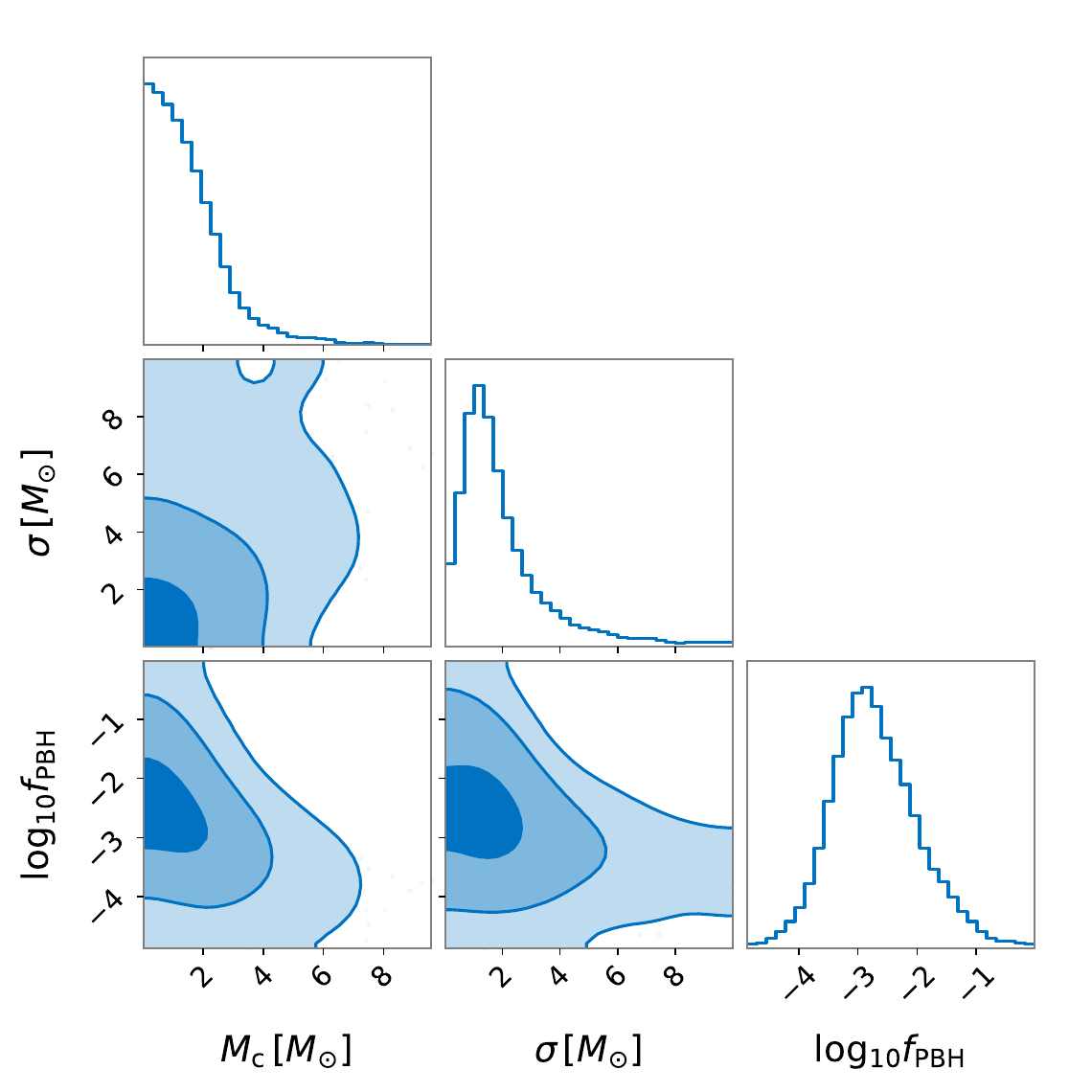}
    \caption{\label{posts_norm} The posterior distribution of the Bayesian parameter estimation for a Gaussian PBH mass function. The contours in the 2D distribution represent $1\sigma$, $2\sigma$ and $3\sigma$ regions respectively.}
\end{figure}  

\begin{figure}[t]
    \centering
    \includegraphics[width=\linewidth]{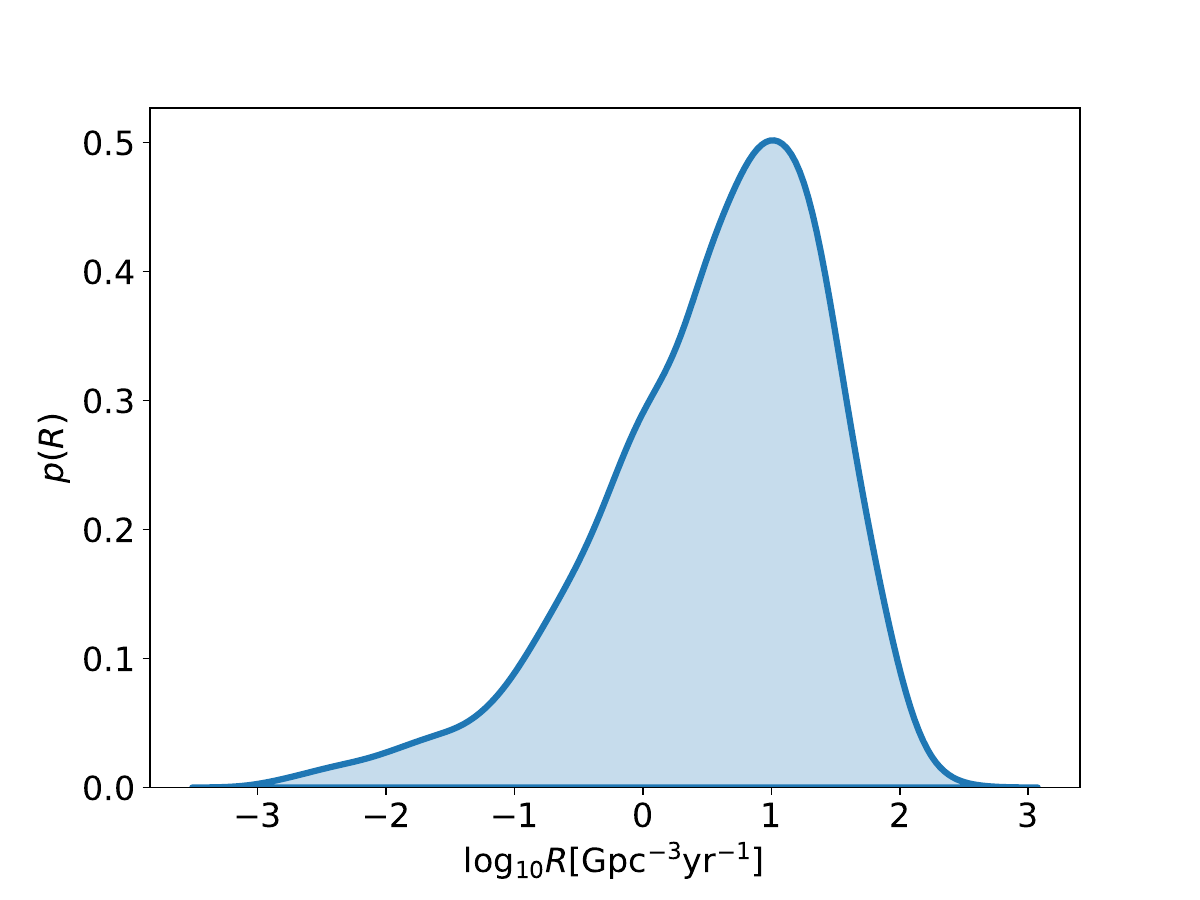}
    \caption{\label{postR_norm}Posterior on the local merger rate of GW230529-like PBH-PBH systems.}
\end{figure}  

\begin{figure}[t]
    \centering
    \includegraphics[width=\linewidth]{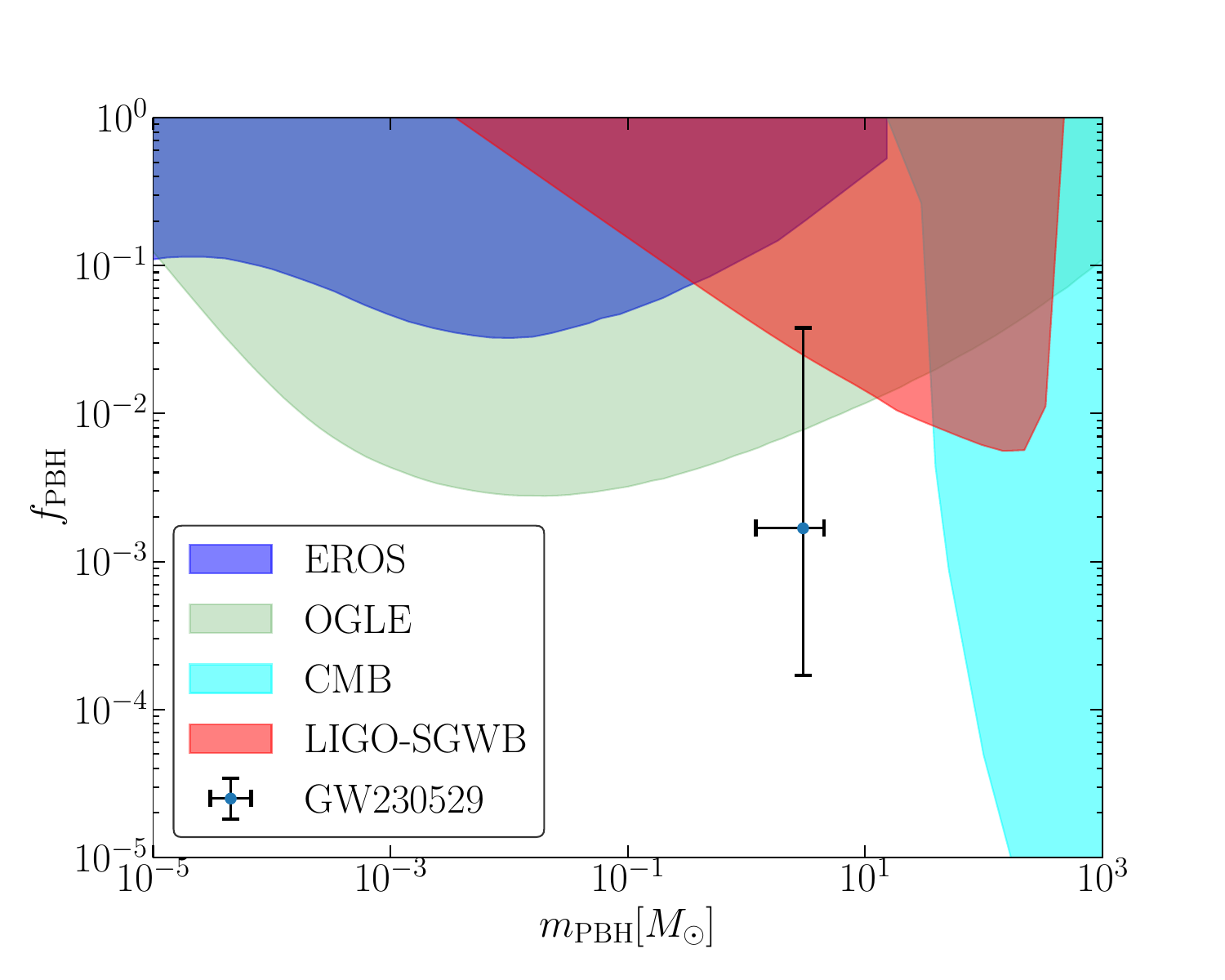}
    \caption{\label{constraint}Constraints on $\fpbh$ with $90\%$ CL error bars inferred from the merger rates for GW230529. The existing upper limits on $\fpbh$ at $90\%$ CL from EROS/MACHO~\cite{EROS-2:2006ryy}, OGLE~\cite{Mroz:2024mse,Mroz:2024wag}, LIGO SGWB~\cite{Wang:2016ana,Chen:2019irf}, and CMB~\cite{Ali-Haimoud:2016mbv,Blum:2016cjs,Horowitz:2016lib,Chen:2016pud,Poulin:2017bwe} are shown for comparison.}
\end{figure}  

In our analysis, we employ uniform priors for each parameter: $M_c$ in the range $[{0}, 10] \, \Msun$, $\sigma$ in the range $[{0}, 10]\, \Msun$, and $\log_{10} \fpbh$ in the range $[-6, 0]$.
We report the Bayesian posterior distribution of parameters for a Gaussian PBH mass function in Fig.~\ref{posts_norm}. The median value of the parameters together with their $90\%$ equal-tailed credible intervals are $\sigma=1.52^{+4.14}_{-0.98}\,\Msun$ and $\fpbh=1.7^{+36.2}_{-1.5}\times 10^{-3}$, while the median mass $M_c$ is constrained to be $M_c \lesssim 3.82\, \Msun$ at a $95\%$ credible level. Moreover, we present the posterior distribution of the local merger rate in Fig.~\ref{postR_norm} whose value is $5.0^{+47.0}_{-4.9} \gpcyr$ derived using \Eq{Rt}.
We also compare our result with current observational constraints on $\fpbh$ in Fig.~\ref{constraint}. The observational constraints include EROS/MACHO microlensing \cite{EROS-2:2006ryy}, OGLE microlensing \cite{Mroz:2024mse,Mroz:2024wag}, stochastic GW background (SGWB) from binary PBHs \cite{Wang:2016ana,Chen:2019irf} and accretion constraints of cosmic microwave background (CMB) \cite{Ali-Haimoud:2016mbv,Blum:2016cjs,Horowitz:2016lib,Chen:2016pud,Poulin:2017bwe}. Although $\fpbh$ has some overlap with the latest OGLE microlensing result and LIGO-SGWB, it is still consistent with current observations. As future data accumulate, it may be possible to verify or falsify the primordial origins of GW230529-like events.

In this work, we investigate the potential origin of GW230529, detected by the LVK.
The identification of the event's component masses as falling within the ``mass gap'' presents a compelling argument against their classification as either NS or ABH. Such a distinction not only challenges existing astrophysical models but also directs our attention toward the possibility of PBHs as a plausible explanation. 
Assuming that GW230529 is indeed a PBH-PBH merger, this event could lead to a fraction, $\fpbh=1.7^{+36.2}_{-1.5} \times 10^{-3}$, of the PBH dark matter. Furthermore, the required PBH abundance is consistent with existing upper limits from microlensing, CMB observations, and SGWB.
The estimated local merger rate is $5.0^{+47.0}_{-4.9} \gpcyr$ for GW230529-like events.
The hypothesis that GW230529 is the result of a PBH-PBH merger is supported by several considerations.
Firstly, the ``mass gap'' evidence reduces the likelihood that the primary component is NS or ABH, given the current understanding of the astrophysical model. Secondly, given that the merger rate of PBH-NS mergers is much lower than a PBH-PBH merger \cite{Sasaki:2021iuc}, this leaves us with the intriguing possibility that we might observe phenomena involving the PBH-PBH merger. 

{It is important to note that our analysis focuses solely on the PBH scenario and does not provide a quantitative comparison with astrophysical models. Our primary motivation for interpreting GW230529 as a PBH event stems from the secondary mass component falling within the lower mass gap. Traditional astrophysical models cannot produce NSs or black holes within this mass range, leading us to assign a prior of zero for these models during our data analysis. Consequently, computing Bayes factors between the primordial and traditional astrophysical models would be infeasible. Moreover, considering modified astrophysical models that might account for the lower mass gap is beyond the scope of our current study. We recognize that a comprehensive assessment would benefit from comparing the PBH interpretation with alternative explanations, and we leave future work to explore this aspect further.}

In conclusion, the analysis of GW230529 presents an intriguing case for the occurrence of a PBH-PBH merger with consistency in the required PBH abundance to represent the dark matter and the local merger rate of PBHs. Further observations are essential to confirm this hypothesis. As we continue to probe the cosmos and with the future data accumulated, it may be possible to verify or falsify the PBH-PBH merger hypothesis.

\emph{Acknowledgments.}
QGH is supported by the National Key Research and Development Program of China Grant No.2020YFC2201502, the grants from NSFC (Grant No.~12250010, 11991052) and Key Research Program of Frontier Sciences, CAS, Grant No.~ZDBS-LY-7009. 
ZCC is supported by the National Natural Science Foundation of China (Grant No.~12247176 and No.~12247112) and the innovative research group of Hunan Province under Grant No. 2024JJ1006.
LL is supported by the National Natural Science Foundation of China (Grant No.~12247112 and No.~12247176) and the China Postdoctoral Science Foundation Fellowship No.~2023M730300.
C.Y. acknowledge the financial support provided under the European Union’s H2020 ERC Advanced Grant “Black holes: gravitational engines of discovery” grant agreement no. Gravitas-101052587. Views and opinions expressed are however those of the author only and do not necessarily reflect those of the European Union or the European Research Council. Neither the European Union nor the granting authority can be held responsible for them. We acknowledge support from the Villum Investigator program supported by the VILLUM Foundation (grant no. VIL37766) and the DNRF Chair program (grant no. DNRF162) by the Danish National Research Foundation.
This project has received funding from the European Union's Horizon 2020 research and innovation programme under the Marie Sklodowska-Curie grant agreements No 101007855 and No 101131233. 

\bibliographystyle{JHEP}
\bibliography{ref}

\providecommand{\href}[2]{#2}\begingroup\raggedright\begin{thebibliography}{10}

\bibitem{Bailyn:1997xt}
C.D.~Bailyn, R.K.~Jain, P.~Coppi and J.A.~Orosz, \emph{{The Mass distribution of stellar black holes}}, \href{https://doi.org/10.1086/305614}{\emph{Astrophys. J.} {\bfseries 499} (1998) 367} [\href{https://arxiv.org/abs/astro-ph/9708032}{{\ttfamily astro-ph/9708032}}].

\bibitem{Ozel:2010su}
F.~Ozel, D.~Psaltis, R.~Narayan and J.E.~McClintock, \emph{{The Black Hole Mass Distribution in the Galaxy}}, \href{https://doi.org/10.1088/0004-637X/725/2/1918}{\emph{Astrophys. J.} {\bfseries 725} (2010) 1918} [\href{https://arxiv.org/abs/1006.2834}{{\ttfamily 1006.2834}}].

\bibitem{Farr:2010tu}
W.M.~Farr, N.~Sravan, A.~Cantrell, L.~Kreidberg, C.D.~Bailyn, I.~Mandel et~al., \emph{{The Mass Distribution of Stellar-Mass Black Holes}}, \href{https://doi.org/10.1088/0004-637X/741/2/103}{\emph{Astrophys. J.} {\bfseries 741} (2011) 103} [\href{https://arxiv.org/abs/1011.1459}{{\ttfamily 1011.1459}}].

\bibitem{Fishbach:2020ryj}
M.~Fishbach, R.~Essick and D.E.~Holz, \emph{{Does Matter Matter? Using the mass distribution to distinguish neutron stars and black holes}}, \href{https://doi.org/10.3847/2041-8213/aba7b6}{\emph{Astrophys. J. Lett.} {\bfseries 899} (2020) L8} [\href{https://arxiv.org/abs/2006.13178}{{\ttfamily 2006.13178}}].

\bibitem{LIGOScientific:2020zkf}
{\scshape LIGO Scientific, Virgo} collaboration, \emph{{GW190814: Gravitational Waves from the Coalescence of a 23 Solar Mass Black Hole with a 2.6 Solar Mass Compact Object}}, \href{https://doi.org/10.3847/2041-8213/ab960f}{\emph{Astrophys. J. Lett.} {\bfseries 896} (2020) L44} [\href{https://arxiv.org/abs/2006.12611}{{\ttfamily 2006.12611}}].

\bibitem{LIGOScientific:2024elc}
{\scshape LIGO Scientific, VIRGO, KAGRA} collaboration, \emph{{Observation of Gravitational Waves from the Coalescence of a $2.5-4.5~M_\odot$ Compact Object and a Neutron Star}},  \href{https://arxiv.org/abs/2404.04248}{{\ttfamily 2404.04248}}.

\bibitem{Zeldovich:1967lct}
Y.B.~Zel'dovich and I.D.~Novikov, \emph{{The Hypothesis of Cores Retarded during Expansion and the Hot Cosmological Model}}, {\emph{Sov. Astron.} {\bfseries 10} (1967) 602}.

\bibitem{Hawking:1971ei}
S.~Hawking, \emph{{Gravitationally collapsed objects of very low mass}}, \href{https://doi.org/10.1093/mnras/152.1.75}{\emph{Mon. Not. Roy. Astron. Soc.} {\bfseries 152} (1971) 75}.

\bibitem{Carr:1974nx}
B.J.~Carr and S.W.~Hawking, \emph{{Black holes in the early Universe}}, \href{https://doi.org/10.1093/mnras/168.2.399}{\emph{Mon. Not. Roy. Astron. Soc.} {\bfseries 168} (1974) 399}.

\bibitem{Chen:2018rzo}
Z.-C.~Chen, F.~Huang and Q.-G.~Huang, \emph{{Stochastic Gravitational-wave Background from Binary Black Holes and Binary Neutron Stars and Implications for LISA}}, \href{https://doi.org/10.3847/1538-4357/aaf581}{\emph{Astrophys. J.} {\bfseries 871} (2019) 97} [\href{https://arxiv.org/abs/1809.10360}{{\ttfamily 1809.10360}}].

\bibitem{Wang:2019kaf}
S.~Wang, T.~Terada and K.~Kohri, \emph{{Prospective constraints on the primordial black hole abundance from the stochastic gravitational-wave backgrounds produced by coalescing events and curvature perturbations}}, \href{https://doi.org/10.1103/PhysRevD.99.103531}{\emph{Phys. Rev. D} {\bfseries 99} (2019) 103531} [\href{https://arxiv.org/abs/1903.05924}{{\ttfamily 1903.05924}}].

\bibitem{Liu:2020cds}
L.~Liu, Z.-K.~Guo, R.-G.~Cai and S.P.~Kim, \emph{{Merger rate distribution of primordial black hole binaries with electric charges}}, \href{https://doi.org/10.1103/PhysRevD.102.043508}{\emph{Phys. Rev. D} {\bfseries 102} (2020) 043508} [\href{https://arxiv.org/abs/2001.02984}{{\ttfamily 2001.02984}}].

\bibitem{DeLuca:2020sae}
V.~De~Luca, V.~Desjacques, G.~Franciolini, P.~Pani and A.~Riotto, \emph{{GW190521 Mass Gap Event and the Primordial Black Hole Scenario}}, \href{https://doi.org/10.1103/PhysRevLett.126.051101}{\emph{Phys. Rev. Lett.} {\bfseries 126} (2021) 051101} [\href{https://arxiv.org/abs/2009.01728}{{\ttfamily 2009.01728}}].

\bibitem{Vaskonen:2020lbd}
V.~Vaskonen and H.~Veerm\"ae, \emph{{Did NANOGrav see a signal from primordial black hole formation?}}, \href{https://doi.org/10.1103/PhysRevLett.126.051303}{\emph{Phys. Rev. Lett.} {\bfseries 126} (2021) 051303} [\href{https://arxiv.org/abs/2009.07832}{{\ttfamily 2009.07832}}].

\bibitem{DeLuca:2020agl}
V.~De~Luca, G.~Franciolini and A.~Riotto, \emph{{NANOGrav Data Hints at Primordial Black Holes as Dark Matter}}, \href{https://doi.org/10.1103/PhysRevLett.126.041303}{\emph{Phys. Rev. Lett.} {\bfseries 126} (2021) 041303} [\href{https://arxiv.org/abs/2009.08268}{{\ttfamily 2009.08268}}].

\bibitem{Domenech:2020ers}
G.~Dom\`enech and S.~Pi, \emph{{NANOGrav hints on planet-mass primordial black holes}}, \href{https://doi.org/10.1007/s11433-021-1839-6}{\emph{Sci. China Phys. Mech. Astron.} {\bfseries 65} (2022) 230411} [\href{https://arxiv.org/abs/2010.03976}{{\ttfamily 2010.03976}}].

\bibitem{Cai:2021wzd}
R.-G.~Cai, C.~Chen and C.~Fu, \emph{{Primordial black holes and stochastic gravitational wave background from inflation with a noncanonical spectator field}}, \href{https://doi.org/10.1103/PhysRevD.104.083537}{\emph{Phys. Rev. D} {\bfseries 104} (2021) 083537} [\href{https://arxiv.org/abs/2108.03422}{{\ttfamily 2108.03422}}].

\bibitem{Yuan:2021qgz}
C.~Yuan and Q.-G.~Huang, \emph{{A topic review on probing primordial black hole dark matter with scalar induced gravitational waves}}, \href{https://doi.org/10.1016/j.isci.2021.102860}{\emph{iScience} {\bfseries 24} (2021) 102860} [\href{https://arxiv.org/abs/2103.04739}{{\ttfamily 2103.04739}}].

\bibitem{Liu:2021jnw}
L.~Liu, X.-Y.~Yang, Z.-K.~Guo and R.-G.~Cai, \emph{{Testing primordial black hole and measuring the Hubble constant with multiband gravitational-wave observations}}, \href{https://doi.org/10.1088/1475-7516/2023/01/006}{\emph{JCAP} {\bfseries 01} (2023) 006} [\href{https://arxiv.org/abs/2112.05473}{{\ttfamily 2112.05473}}].

\bibitem{Liu:2022wtq}
L.~Liu and S.P.~Kim, \emph{{Merger rate of charged black holes from the two-body dynamical capture}}, \href{https://doi.org/10.1088/1475-7516/2022/03/059}{\emph{JCAP} {\bfseries 03} (2022) 059} [\href{https://arxiv.org/abs/2201.02581}{{\ttfamily 2201.02581}}].

\bibitem{Inomata:2022yte}
K.~Inomata, M.~Braglia, X.~Chen and S.~Renaux-Petel, \emph{{Questions on calculation of primordial power spectrum with large spikes: the resonance model case}}, \href{https://doi.org/10.1088/1475-7516/2023/04/011}{\emph{JCAP} {\bfseries 04} (2023) 011} [\href{https://arxiv.org/abs/2211.02586}{{\ttfamily 2211.02586}}].

\bibitem{Meng:2022low}
D.-S.~Meng, C.~Yuan and Q.-G.~Huang, \emph{{Primordial black holes generated by the non-minimal spectator field}}, \href{https://doi.org/10.1007/s11433-022-2095-5}{\emph{Sci. China Phys. Mech. Astron.} {\bfseries 66} (2023) 280411} [\href{https://arxiv.org/abs/2212.03577}{{\ttfamily 2212.03577}}].

\bibitem{Wang:2024vfv}
X.~Wang, Y.-l.~Zhang and M.~Sasaki, \emph{{Enhanced Curvature Perturbation and Primordial Black Hole Formation in Two-stage Inflation with a break}},  \href{https://arxiv.org/abs/2404.02492}{{\ttfamily 2404.02492}}.

\bibitem{Yuan:2024yyo}
C.~Yuan and Q.-G.~Huang, \emph{{Primordial Black Hole Interpretation in Subsolar Mass Gravitational Wave Candidate SSM200308}},  \href{https://arxiv.org/abs/2404.03328}{{\ttfamily 2404.03328}}.

\bibitem{Chen:2024dxh}
Z.-C.~Chen and A.~Hall, \emph{{Confronting primordial black holes with LIGO-Virgo-KAGRA and the Einstein Telescope}},  \href{https://arxiv.org/abs/2402.03934}{{\ttfamily 2402.03934}}.

\bibitem{Hai-LongHuang:2023atg}
H.-L.~Huang, J.-Q.~Jiang and Y.-S.~Piao, \emph{{Merger rate of supermassive primordial black hole binaries}}, \href{https://doi.org/10.1103/PhysRevD.109.063515}{\emph{Phys. Rev. D} {\bfseries 109} (2024) 063515} [\href{https://arxiv.org/abs/2312.00338}{{\ttfamily 2312.00338}}].

\bibitem{Sasaki:2018dmp}
M.~Sasaki, T.~Suyama, T.~Tanaka and S.~Yokoyama, \emph{{Primordial black holes\textemdash{}perspectives in gravitational wave astronomy}}, \href{https://doi.org/10.1088/1361-6382/aaa7b4}{\emph{Class. Quant. Grav.} {\bfseries 35} (2018) 063001} [\href{https://arxiv.org/abs/1801.05235}{{\ttfamily 1801.05235}}].

\bibitem{Carr:2020gox}
B.~Carr, K.~Kohri, Y.~Sendouda and J.~Yokoyama, \emph{{Constraints on primordial black holes}}, \href{https://doi.org/10.1088/1361-6633/ac1e31}{\emph{Rept. Prog. Phys.} {\bfseries 84} (2021) 116902} [\href{https://arxiv.org/abs/2002.12778}{{\ttfamily 2002.12778}}].

\bibitem{Carr:2020xqk}
B.~Carr and F.~Kuhnel, \emph{{Primordial Black Holes as Dark Matter: Recent Developments}}, \href{https://doi.org/10.1146/annurev-nucl-050520-125911}{\emph{Ann. Rev. Nucl. Part. Sci.} {\bfseries 70} (2020) 355} [\href{https://arxiv.org/abs/2006.02838}{{\ttfamily 2006.02838}}].

\bibitem{Bird:2016dcv}
S.~Bird, I.~Cholis, J.B.~Mu\~noz, Y.~Ali-Ha\"\i{}moud, M.~Kamionkowski, E.D.~Kovetz et~al., \emph{{Did LIGO detect dark matter?}}, \href{https://doi.org/10.1103/PhysRevLett.116.201301}{\emph{Phys. Rev. Lett.} {\bfseries 116} (2016) 201301} [\href{https://arxiv.org/abs/1603.00464}{{\ttfamily 1603.00464}}].

\bibitem{Sasaki:2016jop}
M.~Sasaki, T.~Suyama, T.~Tanaka and S.~Yokoyama, \emph{{Primordial Black Hole Scenario for the Gravitational-Wave Event GW150914}}, \href{https://doi.org/10.1103/PhysRevLett.117.061101}{\emph{Phys. Rev. Lett.} {\bfseries 117} (2016) 061101} [\href{https://arxiv.org/abs/1603.08338}{{\ttfamily 1603.08338}}].

\bibitem{Bean:2002kx}
R.~Bean and J.~Magueijo, \emph{{Could supermassive black holes be quintessential primordial black holes?}}, \href{https://doi.org/10.1103/PhysRevD.66.063505}{\emph{Phys. Rev. D} {\bfseries 66} (2002) 063505} [\href{https://arxiv.org/abs/astro-ph/0204486}{{\ttfamily astro-ph/0204486}}].

\bibitem{Kawasaki:2012kn}
M.~Kawasaki, A.~Kusenko and T.T.~Yanagida, \emph{{Primordial seeds of supermassive black holes}}, \href{https://doi.org/10.1016/j.physletb.2012.03.056}{\emph{Phys. Lett. B} {\bfseries 711} (2012) 1} [\href{https://arxiv.org/abs/1202.3848}{{\ttfamily 1202.3848}}].

\bibitem{Nakama:2017xvq}
T.~Nakama, B.~Carr and J.~Silk, \emph{{Limits on primordial black holes from $\mu$ distortions in cosmic microwave background}}, \href{https://doi.org/10.1103/PhysRevD.97.043525}{\emph{Phys. Rev. D} {\bfseries 97} (2018) 043525} [\href{https://arxiv.org/abs/1710.06945}{{\ttfamily 1710.06945}}].

\bibitem{Carr:2018rid}
B.~Carr and J.~Silk, \emph{{Primordial Black Holes as Generators of Cosmic Structures}}, \href{https://doi.org/10.1093/mnras/sty1204}{\emph{Mon. Not. Roy. Astron. Soc.} {\bfseries 478} (2018) 3756} [\href{https://arxiv.org/abs/1801.00672}{{\ttfamily 1801.00672}}].

\bibitem{Barr:2024wwl}
E.D.~Barr et~al., \emph{{A pulsar in a binary with a compact object in the mass gap between neutron stars and black holes}}, \href{https://doi.org/10.1126/science.adg3005}{\emph{Science} {\bfseries 383} (2024) 275} [\href{https://arxiv.org/abs/2401.09872}{{\ttfamily 2401.09872}}].

\bibitem{Chen:2024joj}
Z.-C.~Chen and L.~Liu, \emph{{Is PSR J0514$-$4002E in a PBH-NS binary?}},  \href{https://arxiv.org/abs/2401.12889}{{\ttfamily 2401.12889}}.

\bibitem{Nakamura:1997sm}
T.~Nakamura, M.~Sasaki, T.~Tanaka and K.S.~Thorne, \emph{{Gravitational waves from coalescing black hole MACHO binaries}}, \href{https://doi.org/10.1086/310886}{\emph{Astrophys. J. Lett.} {\bfseries 487} (1997) L139} [\href{https://arxiv.org/abs/astro-ph/9708060}{{\ttfamily astro-ph/9708060}}].

\bibitem{Stasenko:2024pzd}
V.~Stasenko, \emph{{The redshift evolution of primordial black hole merger rate}},  \href{https://arxiv.org/abs/2403.11325}{{\ttfamily 2403.11325}}.

\bibitem{Ali-Haimoud:2017rtz}
Y.~Ali-Ha\"\i{}moud, E.D.~Kovetz and M.~Kamionkowski, \emph{{Merger rate of primordial black-hole binaries}}, \href{https://doi.org/10.1103/PhysRevD.96.123523}{\emph{Phys. Rev. D} {\bfseries 96} (2017) 123523} [\href{https://arxiv.org/abs/1709.06576}{{\ttfamily 1709.06576}}].

\bibitem{Kritos:2020wcl}
K.~Kritos, V.~De~Luca, G.~Franciolini, A.~Kehagias and A.~Riotto, \emph{{The Astro-Primordial Black Hole Merger Rates: a Reappraisal}}, \href{https://doi.org/10.1088/1475-7516/2021/05/039}{\emph{JCAP} {\bfseries 05} (2021) 039} [\href{https://arxiv.org/abs/2012.03585}{{\ttfamily 2012.03585}}].

\bibitem{Chen:2018czv}
Z.-C.~Chen and Q.-G.~Huang, \emph{{Merger Rate Distribution of Primordial-Black-Hole Binaries}}, \href{https://doi.org/10.3847/1538-4357/aad6e2}{\emph{Astrophys. J.} {\bfseries 864} (2018) 61} [\href{https://arxiv.org/abs/1801.10327}{{\ttfamily 1801.10327}}].

\bibitem{Raidal:2018bbj}
M.~Raidal, C.~Spethmann, V.~Vaskonen and H.~Veerm\"ae, \emph{{Formation and Evolution of Primordial Black Hole Binaries in the Early Universe}}, \href{https://doi.org/10.1088/1475-7516/2019/02/018}{\emph{JCAP} {\bfseries 02} (2019) 018} [\href{https://arxiv.org/abs/1812.01930}{{\ttfamily 1812.01930}}].

\bibitem{Liu:2018ess}
L.~Liu, Z.-K.~Guo and R.-G.~Cai, \emph{{Effects of the surrounding primordial black holes on the merger rate of primordial black hole binaries}}, \href{https://doi.org/10.1103/PhysRevD.99.063523}{\emph{Phys. Rev. D} {\bfseries 99} (2019) 063523} [\href{https://arxiv.org/abs/1812.05376}{{\ttfamily 1812.05376}}].

\bibitem{Hutsi:2020sol}
G.~H\"utsi, M.~Raidal, V.~Vaskonen and H.~Veerm\"ae, \emph{{Two populations of LIGO-Virgo black holes}}, \href{https://doi.org/10.1088/1475-7516/2021/03/068}{\emph{JCAP} {\bfseries 03} (2021) 068} [\href{https://arxiv.org/abs/2012.02786}{{\ttfamily 2012.02786}}].

\bibitem{Liu:2019rnx}
L.~Liu, Z.-K.~Guo and R.-G.~Cai, \emph{{Effects of the merger history on the merger rate density of primordial black hole binaries}}, \href{https://doi.org/10.1140/epjc/s10052-019-7227-0}{\emph{Eur. Phys. J. C} {\bfseries 79} (2019) 717} [\href{https://arxiv.org/abs/1901.07672}{{\ttfamily 1901.07672}}].

\bibitem{Wu:2020drm}
Y.~Wu, \emph{{Merger history of primordial black-hole binaries}}, \href{https://doi.org/10.1103/PhysRevD.101.083008}{\emph{Phys. Rev. D} {\bfseries 101} (2020) 083008} [\href{https://arxiv.org/abs/2001.03833}{{\ttfamily 2001.03833}}].

\bibitem{Liu:2022iuf}
L.~Liu, Z.-Q.~You, Y.~Wu and Z.-C.~Chen, \emph{{Constraining the merger history of primordial-black-hole binaries from GWTC-3}}, \href{https://doi.org/10.1103/PhysRevD.107.063035}{\emph{Phys. Rev. D} {\bfseries 107} (2023) 063035} [\href{https://arxiv.org/abs/2210.16094}{{\ttfamily 2210.16094}}].

\bibitem{Planck:2018vyg}
{\scshape Planck} collaboration, \emph{{Planck 2018 results. VI. Cosmological parameters}}, \href{https://doi.org/10.1051/0004-6361/201833910}{\emph{Astron. Astrophys.} {\bfseries 641} (2020) A6} [\href{https://arxiv.org/abs/1807.06209}{{\ttfamily 1807.06209}}].

\bibitem{Dolgov:1992pu}
A.~Dolgov and J.~Silk, \emph{{Baryon isocurvature fluctuations at small scales and baryonic dark matter}}, \href{https://doi.org/10.1103/PhysRevD.47.4244}{\emph{Phys. Rev. D} {\bfseries 47} (1993) 4244}.

\bibitem{Gow:2020cou}
A.D.~Gow, C.T.~Byrnes and A.~Hall, \emph{{Accurate model for the primordial black hole mass distribution from a peak in the power spectrum}}, \href{https://doi.org/10.1103/PhysRevD.105.023503}{\emph{Phys. Rev. D} {\bfseries 105} (2022) 023503} [\href{https://arxiv.org/abs/2009.03204}{{\ttfamily 2009.03204}}].

\bibitem{Ragavendra:2020sop}
H.V.~Ragavendra, P.~Saha, L.~Sriramkumar and J.~Silk, \emph{{Primordial black holes and secondary gravitational waves from ultraslow roll and punctuated inflation}}, \href{https://doi.org/10.1103/PhysRevD.103.083510}{\emph{Phys. Rev. D} {\bfseries 103} (2021) 083510} [\href{https://arxiv.org/abs/2008.12202}{{\ttfamily 2008.12202}}].

\bibitem{Pattison:2021oen}
C.~Pattison, V.~Vennin, D.~Wands and H.~Assadullahi, \emph{{Ultra-slow-roll inflation with quantum diffusion}}, \href{https://doi.org/10.1088/1475-7516/2021/04/080}{\emph{JCAP} {\bfseries 04} (2021) 080} [\href{https://arxiv.org/abs/2101.05741}{{\ttfamily 2101.05741}}].

\bibitem{Palma:2020ejf}
G.A.~Palma, S.~Sypsas and C.~Zenteno, \emph{{Seeding primordial black holes in multifield inflation}}, \href{https://doi.org/10.1103/PhysRevLett.125.121301}{\emph{Phys. Rev. Lett.} {\bfseries 125} (2020) 121301} [\href{https://arxiv.org/abs/2004.06106}{{\ttfamily 2004.06106}}].

\bibitem{Geller:2022nkr}
S.R.~Geller, W.~Qin, E.~McDonough and D.I.~Kaiser, \emph{{Primordial black holes from multifield inflation with nonminimal couplings}}, \href{https://doi.org/10.1103/PhysRevD.106.063535}{\emph{Phys. Rev. D} {\bfseries 106} (2022) 063535} [\href{https://arxiv.org/abs/2205.04471}{{\ttfamily 2205.04471}}].

\bibitem{Loredo:2004nn}
T.J.~Loredo, \emph{{Accounting for source uncertainties in analyses of astronomical survey data}}, \href{https://doi.org/10.1063/1.1835214}{\emph{AIP Conf. Proc.} {\bfseries 735} (2004) 195} [\href{https://arxiv.org/abs/astro-ph/0409387}{{\ttfamily astro-ph/0409387}}].

\bibitem{Thrane:2018qnx}
E.~Thrane and C.~Talbot, \emph{{An introduction to Bayesian inference in gravitational-wave astronomy: parameter estimation, model selection, and hierarchical models}}, \href{https://doi.org/10.1017/pasa.2019.2}{\emph{Publ. Astron. Soc. Austral.} {\bfseries 36} (2019) e010} [\href{https://arxiv.org/abs/1809.02293}{{\ttfamily 1809.02293}}].

\bibitem{Mandel:2018mve}
I.~Mandel, W.M.~Farr and J.R.~Gair, \emph{{Extracting distribution parameters from multiple uncertain observations with selection biases}}, \href{https://doi.org/10.1093/mnras/stz896}{\emph{Mon. Not. Roy. Astron. Soc.} {\bfseries 486} (2019) 1086} [\href{https://arxiv.org/abs/1809.02063}{{\ttfamily 1809.02063}}].

\bibitem{OShaughnessy:2009szr}
R.~O'Shaughnessy, V.~Kalogera and K.~Belczynski, \emph{{Binary Compact Object Coalescence Rates: The Role of Elliptical Galaxies}}, \href{https://doi.org/10.1088/0004-637X/716/1/615}{\emph{Astrophys. J.} {\bfseries 716} (2010) 615} [\href{https://arxiv.org/abs/0908.3635}{{\ttfamily 0908.3635}}].

\bibitem{ligo_scientific_collaboration_and_virgo_2021_5546676}
{\scshape LIGO Scientific, VIRGO, KAGRA} collaboration, \emph{{GWTC-3: Compact Binary Coalescences Observed by LIGO and Virgo During the Second Part of the Third Observing Run — O3 search sensitivity estimates}}, .

\bibitem{KAGRA:2021duu}
{\scshape KAGRA, VIRGO, LIGO Scientific} collaboration, \emph{{Population of Merging Compact Binaries Inferred Using Gravitational Waves through GWTC-3}}, \href{https://doi.org/10.1103/PhysRevX.13.011048}{\emph{Phys. Rev. X} {\bfseries 13} (2023) 011048} [\href{https://arxiv.org/abs/2111.03634}{{\ttfamily 2111.03634}}].

\bibitem{ligo_scientific_collaboration_2024_10845779}
L.S.~Collaboration, V.~Collaboration and K.~Collaboration, \emph{{Observation of Gravitational Waves from the Coalescence of a 2.5-4.5 Msun Compact Object and a Neutron Star --- Data Release}},  Apr., 2024.
\newblock 10.5281/zenodo.10845779.

\bibitem{Mastrogiovanni:2021wsd}
S.~Mastrogiovanni, K.~Leyde, C.~Karathanasis, E.~Chassande-Mottin, D.A.~Steer, J.~Gair et~al., \emph{{On the importance of source population models for gravitational-wave cosmology}}, \href{https://doi.org/10.1103/PhysRevD.104.062009}{\emph{Phys. Rev. D} {\bfseries 104} (2021) 062009} [\href{https://arxiv.org/abs/2103.14663}{{\ttfamily 2103.14663}}].

\bibitem{Speagle:2019ivv}
J.S.~Speagle, \emph{{dynesty: a dynamic nested sampling package for estimating Bayesian posteriors and evidences}}, \href{https://doi.org/10.1093/mnras/staa278}{\emph{Mon. Not. Roy. Astron. Soc.} {\bfseries 493} (2020) 3132} [\href{https://arxiv.org/abs/1904.02180}{{\ttfamily 1904.02180}}].

\bibitem{Ashton:2018jfp}
G.~Ashton et~al., \emph{{BILBY: A user-friendly Bayesian inference library for gravitational-wave astronomy}}, \href{https://doi.org/10.3847/1538-4365/ab06fc}{\emph{Astrophys. J. Suppl.} {\bfseries 241} (2019) 27} [\href{https://arxiv.org/abs/1811.02042}{{\ttfamily 1811.02042}}].

\bibitem{Romero-Shaw:2020owr}
I.M.~Romero-Shaw et~al., \emph{{Bayesian inference for compact binary coalescences with bilby: validation and application to the first LIGO\textendash{}Virgo gravitational-wave transient catalogue}}, \href{https://doi.org/10.1093/mnras/staa2850}{\emph{Mon. Not. Roy. Astron. Soc.} {\bfseries 499} (2020) 3295} [\href{https://arxiv.org/abs/2006.00714}{{\ttfamily 2006.00714}}].

\bibitem{EROS-2:2006ryy}
{\scshape EROS-2} collaboration, \emph{{Limits on the Macho Content of the Galactic Halo from the EROS-2 Survey of the Magellanic Clouds}}, \href{https://doi.org/10.1051/0004-6361:20066017}{\emph{Astron. Astrophys.} {\bfseries 469} (2007) 387} [\href{https://arxiv.org/abs/astro-ph/0607207}{{\ttfamily astro-ph/0607207}}].

\bibitem{Mroz:2024mse}
P.~Mroz et~al., \emph{{No massive black holes in the Milky Way halo}},  \href{https://arxiv.org/abs/2403.02386}{{\ttfamily 2403.02386}}.

\bibitem{Mroz:2024wag}
P.~Mroz et~al., \emph{{Microlensing optical depth and event rate toward the Large Magellanic Cloud based on 20 years of OGLE observations}},  \href{https://arxiv.org/abs/2403.02398}{{\ttfamily 2403.02398}}.

\bibitem{Wang:2016ana}
S.~Wang, Y.-F.~Wang, Q.-G.~Huang and T.G.F.~Li, \emph{{Constraints on the Primordial Black Hole Abundance from the First Advanced LIGO Observation Run Using the Stochastic Gravitational-Wave Background}}, \href{https://doi.org/10.1103/PhysRevLett.120.191102}{\emph{Phys. Rev. Lett.} {\bfseries 120} (2018) 191102} [\href{https://arxiv.org/abs/1610.08725}{{\ttfamily 1610.08725}}].

\bibitem{Chen:2019irf}
Z.-C.~Chen and Q.-G.~Huang, \emph{{Distinguishing Primordial Black Holes from Astrophysical Black Holes by Einstein Telescope and Cosmic Explorer}}, \href{https://doi.org/10.1088/1475-7516/2020/08/039}{\emph{JCAP} {\bfseries 08} (2020) 039} [\href{https://arxiv.org/abs/1904.02396}{{\ttfamily 1904.02396}}].

\bibitem{Ali-Haimoud:2016mbv}
Y.~Ali-Ha\"\i{}moud and M.~Kamionkowski, \emph{{Cosmic microwave background limits on accreting primordial black holes}}, \href{https://doi.org/10.1103/PhysRevD.95.043534}{\emph{Phys. Rev. D} {\bfseries 95} (2017) 043534} [\href{https://arxiv.org/abs/1612.05644}{{\ttfamily 1612.05644}}].

\bibitem{Blum:2016cjs}
D.~Aloni, K.~Blum and R.~Flauger, \emph{{Cosmic microwave background constraints on primordial black hole dark matter}}, \href{https://doi.org/10.1088/1475-7516/2017/05/017}{\emph{JCAP} {\bfseries 05} (2017) 017} [\href{https://arxiv.org/abs/1612.06811}{{\ttfamily 1612.06811}}].

\bibitem{Horowitz:2016lib}
B.~Horowitz, \emph{{Revisiting Primordial Black Holes Constraints from Ionization History}},  \href{https://arxiv.org/abs/1612.07264}{{\ttfamily 1612.07264}}.

\bibitem{Chen:2016pud}
L.~Chen, Q.-G.~Huang and K.~Wang, \emph{{Constraint on the abundance of primordial black holes in dark matter from Planck data}}, \href{https://doi.org/10.1088/1475-7516/2016/12/044}{\emph{JCAP} {\bfseries 12} (2016) 044} [\href{https://arxiv.org/abs/1608.02174}{{\ttfamily 1608.02174}}].

\bibitem{Poulin:2017bwe}
V.~Poulin, P.D.~Serpico, F.~Calore, S.~Clesse and K.~Kohri, \emph{{CMB bounds on disk-accreting massive primordial black holes}}, \href{https://doi.org/10.1103/PhysRevD.96.083524}{\emph{Phys. Rev. D} {\bfseries 96} (2017) 083524} [\href{https://arxiv.org/abs/1707.04206}{{\ttfamily 1707.04206}}].

\bibitem{Sasaki:2021iuc}
M.~Sasaki, V.~Takhistov, V.~Vardanyan and Y.-l.~Zhang, \emph{{Establishing the Nonprimordial Origin of Black Hole\textendash{}Neutron Star Mergers}}, \href{https://doi.org/10.3847/1538-4357/ac66da}{\emph{Astrophys. J.} {\bfseries 931} (2022) 2} [\href{https://arxiv.org/abs/2110.09509}{{\ttfamily 2110.09509}}].

\end{thebibliography}\endgroup
\end{document}